# Controlling the electro-optic response of a semiconducting perovskite coupled to a phonon-resonant cavity

Lucia Di Virgilio[1], Jaco J. Geuchies[1], Heejae Kim[1,2], Keno Krewer[1], Hai Wang[1], Maksim Grechko[1*], Mischa Bonn[1*]

[1] Max Planck Institute for Polymer Research, Ackermannweg 10, 55128 Mainz, Germany.

[2] Pohang University of Science and Technology, Department of Physics, 37673 Pohang, Korea.

*Correspondence to: grechko@mpip-mainz.mpg.de, bonn@mpip-mainz.mpg.de

**Optical cavities, resonant with vibrational or electronic transitions of material within the cavity, enable control of light-matter interaction. Previous studies have reported cavity-induced modifications of chemical reactivity, fluorescence, phase behavior, and charge transport. Here, we explore the effect of resonant cavity-phonon coupling on the transient photoconductivity in a hybrid organic-inorganic perovskite. To this end, we measure the ultrafast photoconductivity response of perovskite in a tunable Fabry–Pérot terahertz cavity, designed to be transparent for optical excitation. The terahertz-cavity field-phonon interaction causes apparent Rabi splitting between the perovskite phonon mode and the cavity mode. We explore whether the cavity-phonon interaction affects the material's electron-phonon interaction by determining the charge carrier mobility through the photoconductivity. Despite the apparent hybridization of cavity and phonon modes, we show that the perovskite properties, in both ground (phonon response) and excited (photoconductive response) states, remain unaffected by the tunable light-matter interaction. Yet the response of the integral perovskite-terahertz optical cavity system depends critically on the interaction strength of the cavity with the phonon: the transient terahertz response to optical excitation can be increased up to 3-fold by tuning the cavity-perovskite interaction strength. These results enable tunable switches and frequency-controlled induced transparency devices.**

**Introduction**

The selective interaction of coherent electromagnetic radiation with specific microscopic material motions (electronic and/or vibrational) can induce superconductivity [1], optical phase transitions [2,3], and influence molecular reaction pathways [4–6]. Exploiting the quantum nature of light has recently gained increasing attention as an alternative approach. In this method, the material behavior is modulated by tuning its interaction with the vacuum – rather than the coherent – state of the electromagnetic field [7,8]. Utilization of the vacuum state allows for avoiding limitations due to energy dissipation. The interaction between the quantum field and a two-level system causes the mixing of states $|i\rangle_m|j\rangle_f$, composed of material ($|i\rangle_m$) and field ($|j\rangle_f$) states, with different population quantum numbers $i$ and $j$. The material-field interaction is typically weak and only slightly modulates material behaviour (for example, the Lamb shift in hydrogen atoms) [9]. To enhance the coupling, the material can be placed inside a cavity that is tuned in resonance with a transition between the material's two energy levels [7,10–12]. Strong mixing of field and material states using this method has been reported to produce polariton states for several systems (inorganic[13], organic[14,15] and hybrid perovskite[16,17] materials) spanning from visible to terahertz (THz) frequencies. [18–21] This approach significantly affects the rate of chemical reactions[22–25], the supramolecular self-assembly[26,27], and the conductivity[28–33].

Here, we investigate the resonant interaction between a terahertz (THz) field and organic-inorganic perovskite in a cavity and how it affects the electrical and optical properties of the system. The mobility in perovskites is significantly influenced by electron-phonon interactions, giving rise to polaron formation and electron-phonon scattering. The latter is the dominant mechanism inhibiting free charge motion in perovskite [34,35]. In harmonic systems, electron-phonon scattering occurs with dark longitudinal (LO) phonon modes [36] that cannot couple with a transverse electromagnetic field. However, substantial anharmonicity in perovskites leads to mixing of longitudinal with bright transverse (TO) phonon modes and even to actually localized character of these vibrations [37,38]. The latter is likely to enable a variety of scattering pathways for electrons by breaking momentum conservation rules. Thus, coupling of bright phonons with an electromagnetic cavity field can affect the density of states and the character of both TO and LO phonons, and their interaction with charge carriers (**Fig.1a**).

We tune the cavity in resonance with phonon modes of methylammonium lead iodide (MAPI) perovskite to enhance their interaction with the terahertz field. The cavity consists of two fused silica substrates, each with a deposited thin (190 nm) ITO layer (**Fig.1b**). The ITO is transparent in the visible spectral range but is electrically conductive and, thus, reflects a major fraction of THz radiation. This design allows us to photoexcite charge carriers in the perovskite coupled with a THz cavity and probe the charge-carrier mobility using a THz pulse. It can provide an opportunity to control conductivity properties of perovskites and the THz response of a perovskite-cavity system.

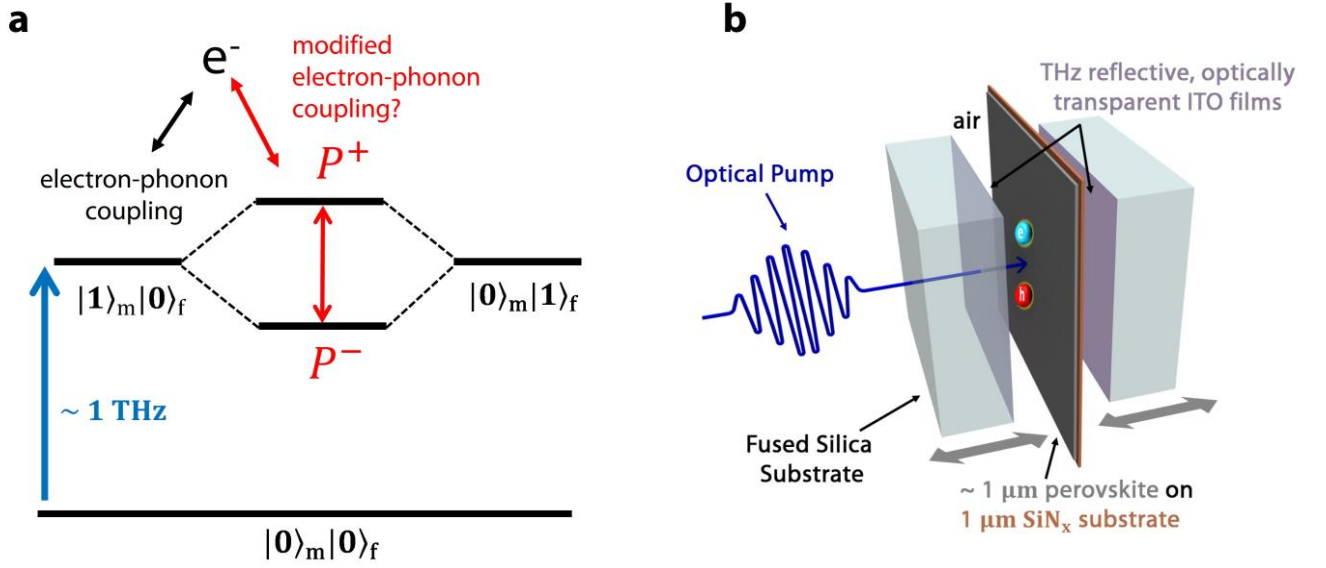

**Figure 1**: **Electron-phonon coupling in an optically accessible, terahertz-tuneable cavity (a)** Schematic representation of the coupling of material and field states producing hybrid light-matter polariton states $P^+$ and $P^-$. We explore here whether the formation of phonon-polaritons can modify electron-phonon interactions. **(b)** Sketch of a perovskite thin film in an optically transparent Fabry–Pérot cavity, with a tunable cavity length of ~100's of μm, so that its fundamental mode resonates at THz frequencies.

## Results

### Optical properties of the perovskite-cavity system

We characterize the performance of the cavity in the THz frequency range by measuring its transmission using terahertz time-domain spectroscopy (THz TDS) [39]. **Figures 2a-f** show the time profile and spectrum of the THz pulse transmitted through the cavity with $x_{gap} = 1$ cm, 345 μm, and 158 μm spacing between the mirrors. For $x_{gap} = 1$ cm, the time delay between consecutive reflections of the THz pulse ($\approx 66$ ps) exceeds the time window of the THz TDS measurement ($\approx 20$ ps). Thus, this measurement characterizes the temporal shape $E_{THz}(t)$ and spectrum $E_{THz}(\omega)$ of the pristine THz pulse (the optical response of the mirrors in this frequency range is shown in Supplementary Fig.1). Upon reducing the cavity length, multiple replicas of the pulse reflecting between the two cavity mirrors appear with spacings of 2.3 ps ($x_{gap} = 345$ μm) and 1.05 ps ($x_{gap} = 158$ μm). The cavity has a quality factor of ~30 (see Supplementary Fig.2), giving rise to a gradually decaying amplitude of the consecutive reflections ($E_{THz}^{345}(t)$ and $E_{THz}^{158}(t)$, Fig. 2b,c). In the frequency domain, the spectra of the original pulse and its reflections interfere, producing characteristic spectra with intensity maxima ($E_{THz}^{345}(\omega)$ and $E_{THz}^{158}(\omega)$, Fig. 2d-f). To derive the transmittance of the cavity, which depends only on the cavity parameters, we divide the intensity spectra $\left|E_{THz}^{345}(\omega)\right|^2$ and $\left|E_{THz}^{158}(\omega)\right|^2$ by the intensity spectrum $|E_{THz}(\omega)|^2$. Depending on the cavity length, the transmittance spectra (Figs. 2d, f) contain few to several peaks, which correspond to the longitudinal modes of the cavity separated by the free spectral range. The smaller the cavity length, the larger its free spectral range, and the sparser the cavity modes. Thus, adjusting $x_{gap}$ allows tuning the cavity modes into resonance with specific optical transitions of a sample inside the cavity.

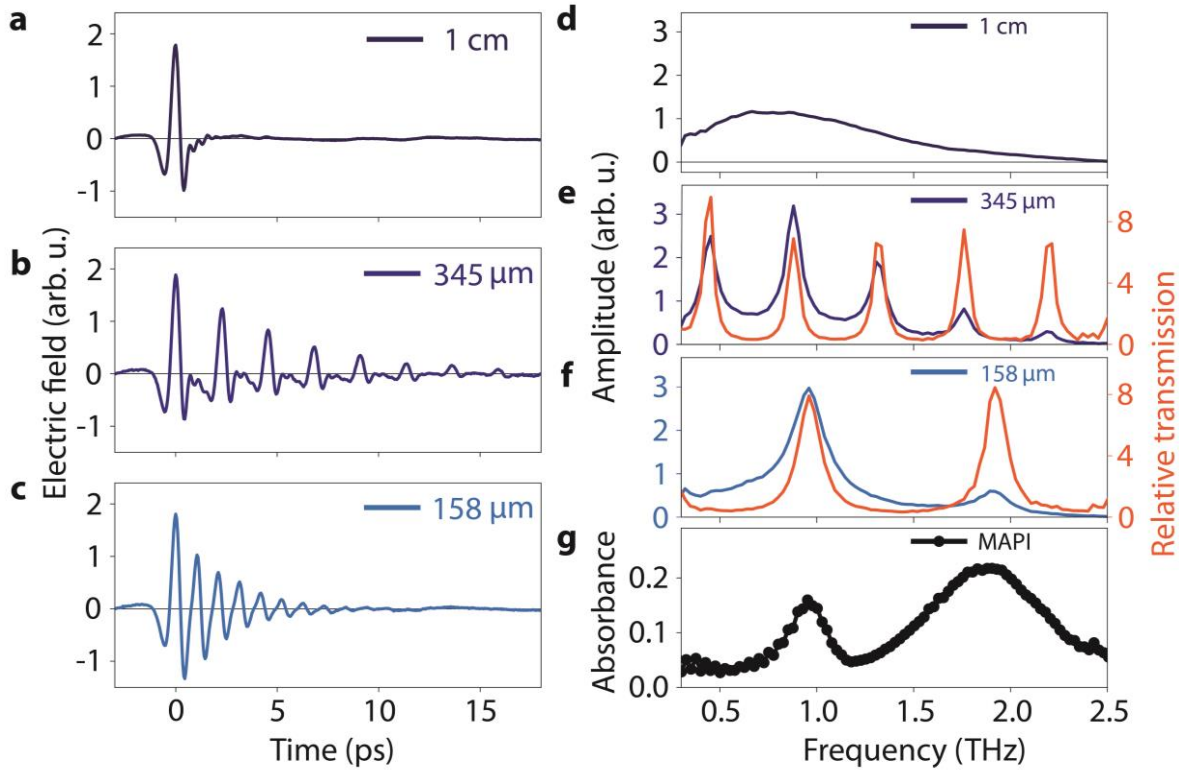

**Figure 2**: **Optical properties of the THz cavity and the perovskite (a)** THz TDS of the empty cavity with a gap of $x_{gap} = 1$ cm; **(b)** $x_{gap} = 345$ µm; **(c)** $x_{gap} = 158$ µm. **(d)** Amplitude of the transmission obtained from the Fourier transform of the time trace in (a). **(e-f)** Frequency-domain transmission spectra for $x_{gap} = 345$ µm and 158 µm. Blue lines show the amplitudes of the transmission; orange lines show the transmission spectrum (amplitude normalized to the spectrum for $x_{gap} = 1$ cm). **(g)** Linear absorbance of methylammonium lead iodide (MAPI) perovskite outside the cavity in the THz frequency range.

We aim to couple these cavity modes with the optically active phonon vibrations of a MAPI perovskite sample. The perovskite has two intense modes in the frequency range of our THz pulse at ~1 THz and ~2 (**Fig. 2g**), which are assigned to the Pb-I-Pb bending vibration and the Pb–I bond stretch [40]. We insert a ~1 µm thick polycrystalline sample supported on 1 µm thick SiNx membrane [41] inside the cavity (see Supplementary Fig.3) and tune the first cavity mode (mode order $m = 1$) close to the resonance with the 1 THz phonon (**Fig. 3a**). With this cavity length, the second cavity mode ($m = 2$) is in resonance with the 2 THz perovskite phonon. However, the interaction of the electromagnetic field of the first and second cavity modes with the thin film sample is different because of the different spatial distributions of the field inside the cavity [42]. In the center of the cavity, the electric field is maximum for the $m = 1$ and zero for the $m = 2$ mode [42]. Thus, the interaction strength (the product of the electric field and the transition dipole moment) for different modes critically depends on the location of the sample inside the cavity. The perovskite-cavity system response shows large variability when the cavity length and/or position of the perovskite within the cavity are varied, as shown in simulations (detailed below) in **Fig. 3b**, and experimentally observed, as shown in **Fig. 3c**. Figure 3b shows that when the sample is near one of the mirrors ($x = 5$ and $138\ \mu m$), polariton formation is suppressed due to the low local field strengths, and the response of the systems closely resembles that of the empty cavity. In between these extrema, the local field strength of the 1 and 2 THz modes determines the field-matter interaction strength. The measurements in Fig. 3c confirm this behavior.

**Photoconductivity of the coupled perovskite-cavity system**

MAPI is a semiconductor, and photoexcitation of its electrons from the valence to the conduction band induces conductivity. We probe conductivity by measuring the change in transmission of a THz pulse with and without optical excitation, in an optical pump-THz probe (OPTP) experiment. To this end, we follow the standard approach and record the relative change $-\Delta E/E$ of the THz electric field at the maximum of THz pulse[43–45]. Optical excitation of MAPI electrons by an ultrafast ($\approx 50$ fs, 400 nm central wavelength) laser pulse induces photoconductivity that decays on a time scale of >100 ps (**Fig. 3d**) [46]. After rapid cooling of the electrons within ~2 ps after photoexcitation, they reach a thermodynamic quasi-equilibrium state [47] with a conductivity almost constant within our 40 ps time window. For identical excitation fluences (for details, see Supplementary Fig.4), we observe indistinguishable charge carrier dynamics for the MAPI sample without a cavity and at different locations inside the cavity (Fig. 3d). Remarkably, the signal at the peak intensity of THz pulse increases by a factor of $\approx 1.5$ when the 1 THz mode of the cavity and perovskite phonon mode are resonant, as opposed to effectively non- (1 cm cavity) and off-resonant cavities. Under the resonant condition, Rabi splitting of 0.25 THz is slightly larger than 0.20 THz and 0.15 widths of the MAPI phonon and cavity modes, respectively. With these parameters, the perovskite-cavity system is in the intermediate (frequently also called strong [31,48–50]) coupling regime [51,52]. It is tempting to assign this increase in transient THz signal to enhanced charge mobility in perovskite, produced by the coupling between the electromagnetic field and the phonon mode. To test this hypothesis, we measure the photoconductivity spectrum 30 ps after photoexcitation, when the sample reaches a quasi-steady state. The spectra look drastically different for the two perovskite-cavity configurations (**Fig. 3e**) across the entire frequency range of the THz pulse. Apparently, their complicated lineshapes show non-uniform change of amplitude and phase of THz pulse caused by photoexcitation. It is challenging to comprehend these results intuitively. Therefore, we perform classical electrodynamics calculations, using the transmission matrix (T-matrix) formalism [53], in order to understand their physical origin.

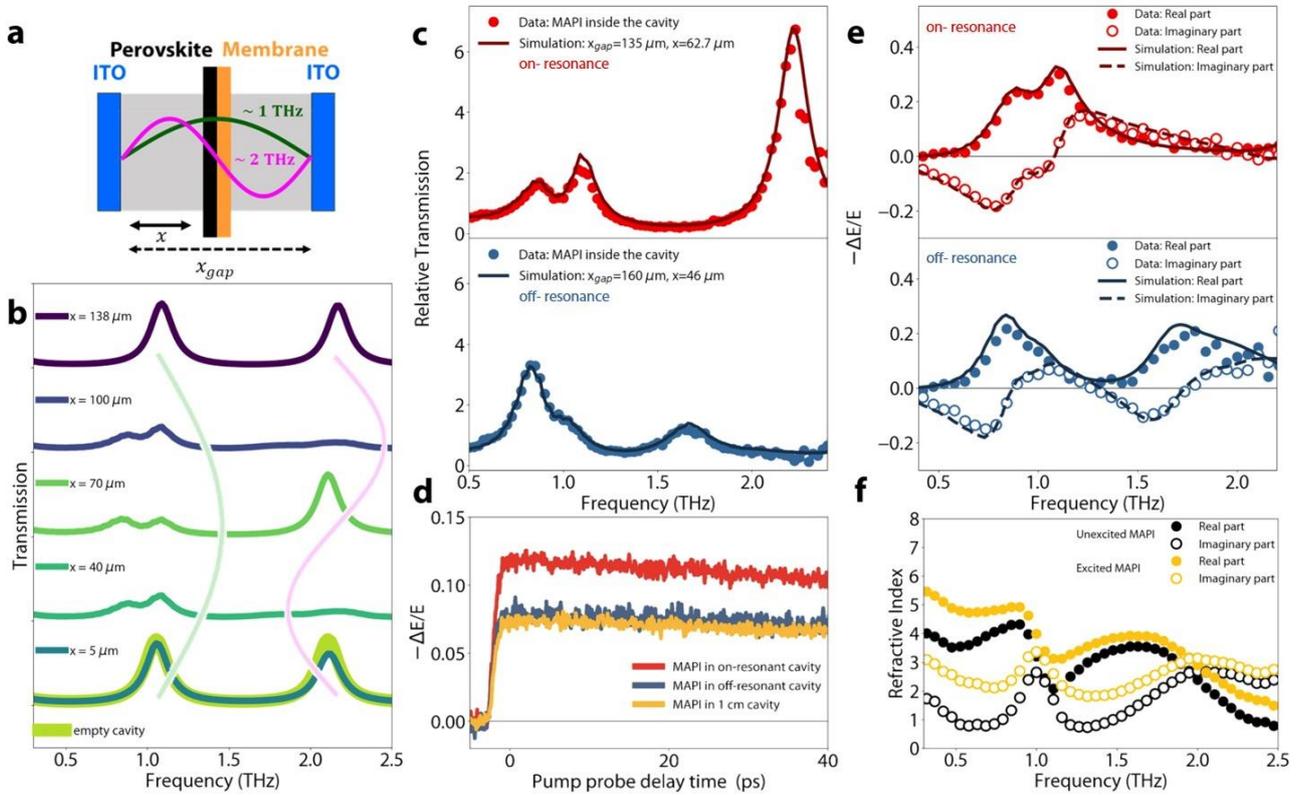

**Figure 3: Frequency- and time-resolved spectra of the perovskite-cavity system (a)** Schematic of the perovskite-cavity system, with the cavity longitudinal $m = 1$ mode at ~1 THz. The field amplitudes of the $m = 1$ and 2 modes (with ~1 and ~2 THz frequency) appear in green and pink. **(b)** Simulations of the spectral transmission of the cavity-perovskite system as a function of the sample position in the cavity, with experimentally determined cavity and perovskite parameters. Depending on the position of the perovskite in the cavity, the local field strength (shaded lines) causes polariton formation at ~1 and ~2 THz **(c)** Experimental (circles) and simulated transmission spectra, for $x_{gap} = 135$ μm and $x = 63$ μm (red line), $x_{gap} = 160$ μm and $x = 46$ μm (blue line). **(d)** Optical pump-THz probe signal of MAPI in the cavity with the same parameters as in (c), and also off-resonant cavity ($x_{gap} = 1$ cm, yellow line). Note the enhanced photoconductivity signal for the on-resonant case (red). **(e)** Experimental (circles) and simulated (thin lines) differential photoconductivity response for the same parameters as in (c). **(f)** Refractive index of (photoexcited) MAPI outside the cavity, used to calculate the response in (c) and (e). All optical pump-THz probe measurements (inside as well as outside the cavity) are performed with an identical pump fluence of around $1 \times 10^{18}$ photons/m$^2$.

**Classical electrodynamics simulations of the (photoexcited) cavity-perovskite system**

Our system, the cavity in resonance with the 1 THz phonon mode of MAPI, is composed of two reflecting ITO surfaces (refractive index independently determined, shown in Supplementary Fig.1) separated by $x_{gap} = 142$ µm, a 1 µm thick perovskite sample with the experimentally determined refractive index shown in **Fig. 3f** (details in Supplementary Note 3), and a 1 µm thick SiNx substrate with a dispersionless refractive index of 2.8 [54]. The calculated spectrum of the transmitted intensity (square of the electric field) is shown in Fig. 3b for various $x$. For $x = 70$ µm the sample is near the node of the $m = 2$ cavity mode and interacts weakly with its field. Thus, the transmission spectrum around 2 THz resembles that of an empty cavity (Fig 3b). In contrast, interaction with the $m = 1$ mode is strong, substantially modulating the transmission spectrum around 1 THz and splitting it into two peaks. This modulation results from the joint action of the imaginary ($k(\omega)$) and real ($n(\omega)$) parts of the sample refractive index. The former causes attenuation of the electromagnetic wave, and the latter changes the interference of the light propagating inside the cavity, by changing the optical path length $n(\omega)x_{gap}$ [55]. For $x = 40$ µm and $100$ µm, the sample interacts substantially with both cavity modes. At the extremes ($x = 5$ µm and $138$ µm), interaction is weak due to the low electric field. These results show that interaction between field and a thin film sample is highly sensitive to the details of the configuration of the sample-cavity system, i.e. $x_{gap}$, $x$ and the sample properties.

Using this model of the system, we reproduce the experimental data by fine-tuning $x_{gap}$ and $x$. The calculation results shown in Fig. 3c are in excellent agreement with the experiment. These calculations demonstrate that, independent of the cavity length and sample location, the THz refractive index of the perovskite measured outside the cavity (see details in Supplementary Note 3 and Supplementary Fig.5) is sufficient to reproduce the transmission spectra of the perovskite-cavity system. However, we observe that the entire system – i.e., the composite of the MAPI and the cavity – after photoexcitation changes its spectral response drastically as a function of both the cavity length and the position of the material, enabled by the tunable light-matter interaction.

To test if these spectra can likewise be explained by classical electrodynamics, we measure the refractive index of MAPI in free space (see Supplementary Fig.6), at the same 30 ps time delay after photoexcitation, under identical excitation conditions (Fig. 3f). Note that presence of the cavity cannot cause interference effects for the pump pulse because the ITO coating is transparent at this wavelength. The thin lines in Fig. 3e show the photoconductivity spectra for the two configurations of the perovskite-cavity system calculated using T-matrix formalism and refractive index of ground-state and excited MAPI measured outside the cavity. This parameter-free description captures the data very well.

This excellent agreement demonstrates that classical electrodynamics can fully account for the complex, non-intuitive change of the perovskite transmission induced by its photoexcitation inside the cavity. In this model, the physical properties of MAPI do not change inside the cavity. Instead, the spectral lineshapes and photoconductivity dynamics are generated by wave absorption and interference, determined in a non-trivial manner by the cavity properties and the perovskite extinction coefficient and refractive index - different in the ground and excited states[55,56]. The change of the field modulation at the maximum of THz pulse in Fig. 3d is the result of complex interference of spectral components shown in Fig. 3e, with their different intensities and phases for different perovskite-cavity configurations.

**On-demand control of transient THz electric field**

Even though the perovskite properties, both in the ground and excited states, do not change significantly inside the terahertz cavity, the response of the entire perovskite-cavity system shows substantial variations. The results in Fig. 3d demonstrate that the system enables tunable control of THz field modulation. These data represent the change of the maximum field of the THz pulse, as discussed above. Yet, modulation is not restricted to the maximum only. We now closely examine the data in the time domain, i. e. before Fourier transformation that produced the spectra in Fig. 3e. For the two system configurations, **Figure 4a** shows the electric field $E(t)$ of the THz pulse and its photoinduced change $-\Delta E(t)$ measured 30 ps after sample excitation. The modulation depth of the THz field $-\Delta E(t)/E(t)$ shown in Figs. 4b-d for different time intervals, varies substantially across the time profile of the THz pulse, as well as with cavity parameters. Before the maximum of the THz pulse, modulation of its field is identically weak for the on- and off-resonant perovskite-cavity configurations. Around the pulse maximum, the field changes by ~5-10% in the on-resonant perovskite-cavity system (**Fig. 4b**). At a later time of about 0.5 ps modulation increases up to 20 % **(Fig. 4c)**, and reaches 40% at about 1.5 ps after the pulse maximum **(Fig. 4d)**. Tuning the cavity into resonance with the 1 THz perovskite mode increases the modulation 1.5, 2 and 3-fold within duration of THz pulse, as shown in **Figs. 4b-d**. Such on-demand adjustability of ultrafast THz field modulation can benefit photonic integrated devices [57] and optical communications modulation [58].

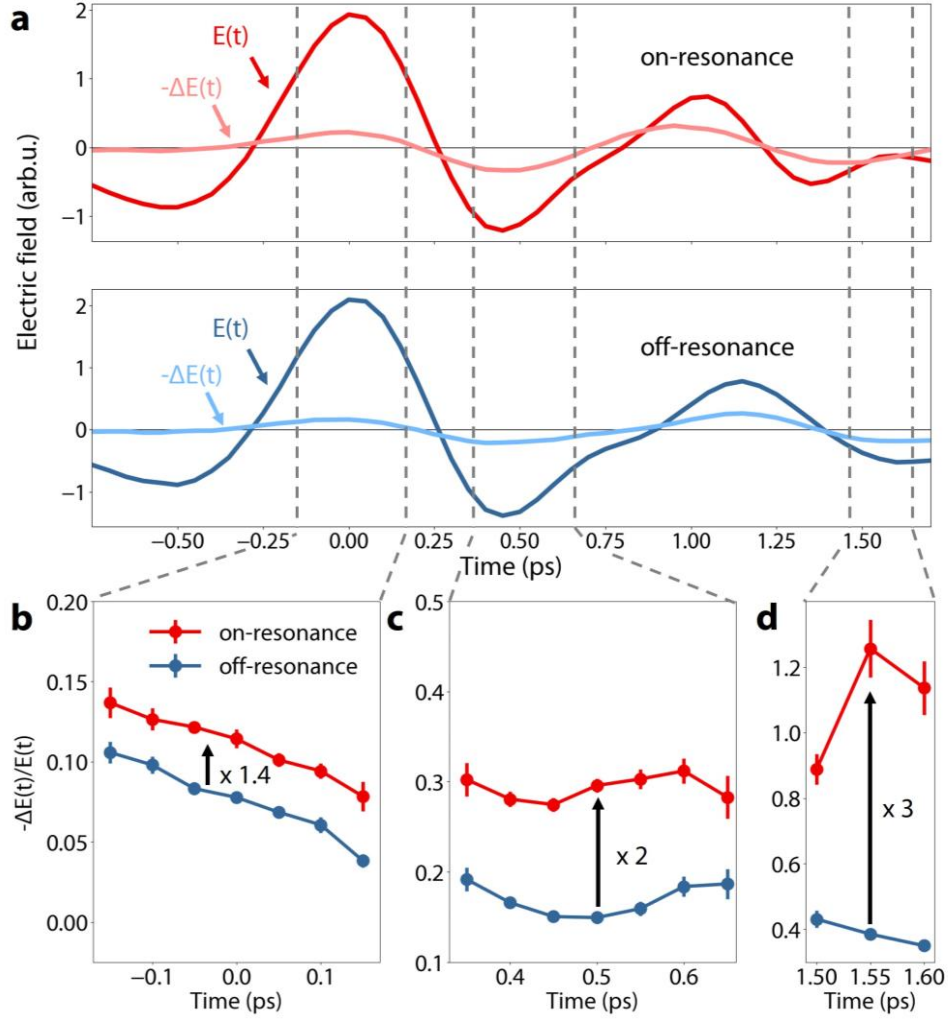

**Figure 4: Cavity-tunable photoconductivity enhancement of the perovskite-cavity system. (a)** Lines without symbols: Time trace of the pulsed THz field $E(t)$ transmitted through the filled cavity on-resonance (red, $x_{gap} = 135$ μm and $x = 62.7$ μm) and off-resonance (blue, $x_{gap} = 160$ μm and $x = 46$ μm). Lines with dots: Relative change in the THz time profile ($-\Delta E(t)/E(t)$) for both configurations. **(b-d)** zoom-in of the data around 0, 0.5, and 1.5 ps highlighted in panel a, showing the relative signal enhancement for the resonant cavity.

**Discussion**

In summary, we report the observation and simulation of the optoelectronic response of a semiconducting perovskite in an optically transparent Fabry–Pérot THz cavity using optical pump-terahertz probe spectroscopy. We reveal a tunable and on-demand control of the THz field modulation through the photoexcited perovskite-cavity system by tuning the cavity-material resonance. Our calculations demonstrate that, despite the apparent formation of polaritons between the cavity and perovskite modes, the material properties of the MAPI in the cavity remain unchanged. However, we observe that the entire MAPI-cavity system changes its spectral response as a function of the cavity length and the position of the material inside the cavity, due to a tunable interaction between the refractive index of the material and the cavity resonances. The measured spectra of both unexcited and photoexcited MAPI inside the cavity can be very well described using classical electrodynamics. This work elucidates the role of cavity resonances in the presence of photoexcited media and opens the possibility of using a transparent THz optical cavity to shape the transmission of THz radiation and enhance, on-demand, THz field modulation through photoexcited semiconductor-cavity systems.

## Methods

### Fabrication of THz cavity

To investigate the coupling between MAPI and cavity resonances, we designed and fabricated a tunable optically transparent THz cavity. The geometry of our system is schematically shown in Fig.1a. The cavity is composed of two ITO films prepared by sputtering around 190 nm ITO layers on 2 mm-thick fused silica purchased from PI-KEM. The two ITO layers are reflective for THz radiation (ITO mirrors). They are mounted on two mirror mounts that allow fine alignment of the cavity. Both ITO mirrors are installed on two translation stages that ensure that allow varying the cavity length continuously and symmetrically around the THz focus (see Supplementary Fig.3). The lateral size of the thin film MAPI (10 mm × 10 mm) is bigger than the size of mirrors (7 mm × 7mm), as schematically shown in Fig.1a.

### Perovskite film preparation

The thin MAPI film was prepared by spin-coating a solution composed of Methylammonium Iodide and PbI into Dimethylformamide on a thin SiNx membrane. Methylammonium iodide (99,9% w/w, MAI) was purchased from Greatcell Energy. Lead acetate (98% w/w, $Pb(Ac)_2$) was purchased from Tokyo Chemical Industry. Anhydrous dimethylformamide (> 99.8% v/v, DMF) was purchased from Sigma Aldrich. SiNx (1 $\mu$m membrane thickness purchased from Norcada) membranes were subjected to UV-Ozone treatment (FHR UVOH 150 LAB, 250 W) for 20 minutes with an oxygen feeding rate of 1L/min right before spin-coating. The substrates were positioned in the center of a microscope slide and held in place with four pieces of electrical tape. Film preparation was carried out in a nitrogen-purged glovebox. The precursor solution consists of 477mg (3mmol) MAI and 325.3mg of $Pb(Ac)_2$ (1mmol) in 1 mL DMF. The film preparation was done by depositing 50 $\mu$L of the precursor solution onto the SiNx membrane and spincoating it at 3000 rpm (ramp ± 1000 rpm/s) for one minute. Afterward, the film was left to dry at room temperature for 5 minutes and subsequently annealed for 10 minutes on a 100°C hotplate. The perovskite-coated membranes were mounted in custom-made holders and stored in the glovebox prior to their use (see more details about MAPI characterization in Supplementary Fig.7)

**Optical Pump THz Probe Spectroscopy**

We use an amplified Ti:sapphire laser producing pulses with 800 nm central wavelength and ~50 fs pulse duration at 1 kHz repetition rate. The THz field is generated by optical rectification in a ZnTe crystal (thickness 1 mm, <110> orientation). The THz detection is based on the electro-optic effect in a second ZnTe crystal with 1 mm thickness. We vary the time delay between THz field and 800 nm sampling beam with a motorized delay stage (M-605.2DD purchased from Physik Instrument (PI)). More details of the THz setup are provided in Ref [59]. The time delay between optical pump and THz probe pulses is controlled by a second motorized delay stage (M-521.DD, Physik Instrument (PI)). The time step in OPTP measurements is 0.1 ps. The pump pulse has 400 nm central wavelength, and is produced by second harmonic generation of 800 nm femtosecond pulse in a beta barium borate crystal (BBO). We performed all measurements with incident photon density $1 \times 10^{18}\ photons/m^2$. All frequency resolved spectra are collected at a fixed pump-probe time delay (30 ps after photoexcitation) by scanning sampling line[60].

## Acknowledgements

This work has received funding from the European Union's Horizon 2020 research and innovation program under the Marie Sklodowska-Curie grant No 811284. L.D.V would like to acknowledge Wenhao Zheng, Shuai Fu and Heng Zhang for useful discussions. We are grateful to Marc-Jan van Zadel for designing and fabricating the MAPI sample holder and excellent technical support. J.J.G. gratefully acknowledges support from the Alexander von Humboldt Foundation.

## Author contributions

M.B. designed the project. M.G. and M.B. supervised the project. L.D.V. conducted the terahertz spectroscopy experiments and developed data analysis tools together with J.J.G.. L.D.V., H.K. and J.J.G. wrote the code to perform the simulations. J.J.G. performed the sample fabrication. All authors contributed to the data analysis, interpretation, and manuscript writing.

## Competing financial interests

The authors declare no competing interests.

## Additional information

Supplementary information accompanies this paper.
Correspondence and requests for materials should be addressed to M.G. (grechko@mpip-mainz.mpg.de) and M.B. (bonn@mpip-mainz.mpg.de).

## Data availability

The data that supports the findings of this study are available from the corresponding authors upon reasonable request.